\documentclass[pra,twocolumn,preprintnumbers,amsmath,amssymb,superscriptaddress]{revtex4-2}

\usepackage[normalem]{ulem}   

\usepackage[utf8]{inputenc}
\usepackage[T1]{fontenc}
\usepackage{mathptmx}
\usepackage{etoolbox}
\usepackage{color}
\usepackage{hyperref}
\usepackage{ulem}
\usepackage{xcolor}
\usepackage{mathrsfs}
\usepackage{physics}

\PassOptionsToPackage{hyphens}{url} 
\usepackage{xurl}                   

\usepackage{microtype}              
\usepackage{soul}

\AtBeginDocument{\RenewCommandCopy\qty\SI}

\usepackage{relsize}
\usepackage{placeins}
\usepackage[T1]{fontenc}
\usepackage{amsmath}
\usepackage{array}
\usepackage{placeins}
\usepackage{siunitx}
\usepackage{booktabs}
\usepackage{rotating}
\usepackage[english]{babel}

\begin{document}


\title{Macroscopic entanglement distribution with atomic ensembles}

\author{Shuang Li}
\thanks{The indicated authors are joint first authors}
\affiliation{State Key Laboratory of Precision Spectroscopy, School of Physical and Material Sciences, East China Normal University, Shanghai 200062, China}

\author{Jin Hu}
\thanks{The indicated authors are joint first authors}
\affiliation{State Key Laboratory of Precision Spectroscopy, School of Physical and Material Sciences, East China Normal University, Shanghai 200062, China}

\author{Ilia D. Lazarev}
\thanks{The indicated authors are joint first authors}
\affiliation{Federal Research Center of Problems of Chemical Physics and Medicinal Chemistry RAS, Acad. Semenov av. 1, Chernogolovka, Moscow region, Russia, 142432}
\affiliation{Faculty of Fundamental Physical-Chemical Engineering, Lomonosov Moscow State University, GSP-1, Moscow, 119991, Russia}

 \author{Jonathan Raghoonanan}
 \affiliation{New York University Shanghai, 567 West Yangsi Road, Shanghai, 200124, China.}
 \affiliation{Department of Physics, New York University, New York, NY 10003, USA}

\author{Valentin Ivannikov}
\email{valentin@nyu.edu}
\affiliation{New York University Shanghai, 567 West Yangsi Road, Shanghai, 200124, China.}

\author{Alexey N. Pyrkov}
\affiliation{Federal Research Center of Problems of Chemical Physics and Medicinal Chemistry RAS, Acad. Semenov av. 1, Chernogolovka, Moscow region, Russia, 142432}

\author{Tim Byrnes}
\email{tim.byrnes@nyu.edu}
\affiliation{New York University Shanghai; NYU-ECNU Institute of Physics at NYU Shanghai, 567 West Yangsi Road, Shanghai, 200124, China.}
\affiliation{State Key Laboratory of Precision Spectroscopy, School of Physical and Material Sciences, East China Normal University, Shanghai 200062, China}
\affiliation{Center for Quantum and Topological Systems (CQTS), NYUAD Research Institute, New York University Abu Dhabi, UAE}
\affiliation{Department of Physics, New York University, New York, NY 10003, USA}

\date{\today}

\begin{abstract}
The distribution of entanglement is a crucial task for quantum communication towards realizing a globe-spanning quantum internet. Recently a protocol for deterministic long-distance distribution of macroscopic entanglement over a network of ensembles of qubits was introduced [Adv.~Quantum~Technol.~\textbf{2025}, 8,~2400524]. It was shown that this protocol allows for the propagation of macroscopic amounts of entanglement with a protocol complexity that is independent on the ensemble size. However, questions remained on whether the scheme is viable, particularly for a large particle number, which is the case for realistic atomic ensembles.   
Here we develop improved numerical techniques that allow calculation of realistic ensemble sizes up to $ \sim 10^6 $ with a negligible loss of accuracy.  We find that moderate dephasing leaves the entanglement largely intact at the magic times, whereas stronger noise monotonically suppresses the entanglement. Our results demonstrate that the protocol retains its functionality towards the macroscopic regime and provides quantitative benchmarks for its robustness under a realistic level of decoherence.
\end{abstract}

\maketitle


\section{\label{sec:level1}Introduction}

One of the major aims in the field of quantum information science is to establish a globe-spanning quantum internet \cite{kimble2008quantum,wehner2018quantum,rohde2025quantum,ruskuc2025multiplexed, hermans2022qubit, simon2017towards,pittaluga2025long, zhou2023experimental,chen2021integrated}, for applications such as distributed quantum computing \cite{main2025distributed, cuomo2023optimized}, entanglement-based quantum cryptography \cite{ekert1991,bennett1992,acin2007}, and quantum metrology and distributed sensing \cite{fadel2023multiparameter,giovannetti2011,degen2017,komar2014}. Most of the quantum protocols for quantum information transmission and processing are based on quantum correlations, especially on quantum entanglement. Photons are the natural technology suited for distributing long-distance entanglement. Demonstrations using space-based distributions have shown it is possible to distribute entanglement in the range of 1200 km \cite{yin2017satellite}. 
For longer distances, the Earth's curvature becomes a fundamental limit for creating entanglement between any two points on Earth.  To enable long-distance entanglement distribution, the quantum repeater protocol was proposed \cite{briegel1998quantum, shi2025measurement, azuma2023quantum}. The protocol is based on entanglement distribution between distant nodes with entanglement swapping, by the use of quantum memories \cite{pu2021experimental} and Bell-state measurements \cite{williams2017superdense}.  Quantum memories are required because the generation of entanglement (for example by parametric down conversion) across the elementary links has probabilistic nature, meaning that the entanglement may not be presented synchronously. A quantum memory stores the entanglement until the entanglement swapping step is ready to be performed \cite{heshami2016quantum,jiang2009quantum}. This allows the extension of the capabilities of quantum networks to arbitrary topologies and distances by dividing long segments into several short-distance links, where entanglement can be directly generated.

There are numerous physical systems which are potential candidates for quantum memories, such as single neutral atoms \cite{rosenfeld2008towards}, atomic ensembles \cite{sangouard2011quantum}, trapped ions \cite{duan2010colloquium}, and the nitrogen-vacancy centers in diamonds \cite{hermans2022qubit}, which can absorb and store incoming photons, for instance using techniques such as electromagnetically induced transparency \cite{fleischhauer2005electromagnetically}. In particular, as a platform to realize a quantum memory, atomic gas ensembles are an attractive candidate due to their ability to perform strong and controllable coupling between the atoms and photons.  This allows to serve such systems as the nodes of a quantum network with an embedded quantum memory \cite{duan2010colloquium, ritter2012elementary}.  Another advantage of using an ensemble is that there is a collective enhancement owing to the large number of atoms with long coherence times \cite{brion2007quantum,lukin2001dipole,byrnes2011accelerated}.  In such schemes using collective enhancement, typically states such as Dicke states are used as discrete levels to couple with single excitations.  Thus, while an atomic ensemble contains an extremely large number of physical atoms and degrees of freedom, only specific energy levels are used. This means that the amount of entanglement that is distributed is at the one ebit level.  

However, the large number of degrees of freedom that are available in the atomic ensemble allow for the possibility of generating {\it macroscopic} entanglement.  In this case, the atomic ensemble acts as a high dimensional qudit, and the entanglement is shared between two or more such qudits.  The nature and applications of such macroscopic entanglement has been studied in works such as Refs. \cite{byrnes2012macroscopic,byrnes2013fractality,byrnes2015macroscopic,chaudhary2021remote,abdelrahman2014coherent}.  Recently, a new quantum repeater protocol to create long-distance macroscopic entanglement between cold atomic ensembles was proposed \cite{pyrkov2025quantum}.  The protocol generalizes the qubit-based quantum repeater protocol to the qudit case with $ S_z S_z $ interactions, and showed that macroscopic entanglement can be efficiently distributed. Here, by efficiently, we mean that the number of steps in the protocol is {\it independent} of the amount of entanglement that is distributed.  However, such macroscopic entanglement can be  highly susceptible to decoherence, in the same way that Schrodinger cat states are extremely fragile.  Questions remain of whether the scheme can be implemented in realistic situations.



In this work, we revisit the protocol introduced in Ref. \cite{pyrkov2025quantum} and develop more advanced numerical techniques that can handle more realistic ensemble sizes.  
For example, in atomic ensembles trapped on an atom chip, the number of atoms can be $10^4$ or more \cite{byrnes2015macroscopic,bohi2009coherent}. In Ref. \cite{pyrkov2025quantum}, the largest system including decoherence was for 3 atoms per ensemble, which is clearly far from a realistic regime. Using our new methods we may more definitively answer whether the scheme still generate entanglement reliably even with a more realistic number of qubits.  
We show how to explore ensemble sizes up to $N=10^4$, mapping the scaling of the end-to-end entanglement with the particle number.  To make such regimes feasible, we employ a Fock space truncation window approximation \cite{karmanov2008systematic} that preserves the relevant physics of the collective Dicke manifold while enabling efficient simulation of the full time evolution and entanglement dynamics. 
We study the system under two conditions, decoherence-free and dephasing, pushing the simulation limits to $N = 10^6$ and $N=30$, respectively.

\section{Quantum repeater for distributing macroscopic entanglement}

\label{sec2}

\subsection{Physical system}

Following Ref. \cite{pyrkov2025quantum}, we consider a linear chain of $M$ ensembles, each holding $N$
qubits.  We assume that the only controls and measurements that are available are through collective spin operators on each ensemble
\begin{equation}
S_\gamma=\sum_{\ell=1}^{N}\sigma^\gamma_\ell,\qquad \gamma\in\{x,y,z\}.
\end{equation}
Assuming the initial state is in the completely symmetric sector, we may use the Schwinger representation with modes \cite{byrnes2021quantum}
\begin{subequations}\label{eq:schwinger} %
\begin{align}
S_x   &= a^\dagger b + b^\dagger a,              \label{eq:schwinger:a}\\
S_y   &= -i\,a^\dagger b + i\,b^\dagger a,       \label{eq:schwinger:b}\\
S_z   &= n_a - n_b,                               \label{eq:schwinger:c}\\
\hat N&= n_a + n_b    .                          \label{eq:schwinger:d}
\end{align}
\end{subequations}
where we defined 
\begin{subequations}\label{eq:init_sc}
\begin{align}
n_a & = a^\dagger a  \\
n_b & = b^\dagger b  .
\end{align}
\end{subequations}
The number (Dicke) states in the $ z $-basis are defined as
\begin{align}
    \ket{k} &= \frac{(a^\dagger)^k (b^\dagger)^{N-k}}{\sqrt{k!\,(N-k)!}}\ket{\varnothing},
\label{eq:dicke:a}
\end{align}
where $ \ket{\varnothing} $ is the vacuum state and the Dicke states satisfy
\begin{subequations}\label{eq:dicke}
\begin{align}
n_a \ket{k} &= k \ket{k} \\
n_b \ket{k} &= (N-k) \ket{k} \\
S_z\ket{k} &= (2k-N)\ket{k} \label{eq:dicke:b} 
\end{align}
\end{subequations}
for $k=0,1,\ldots,N$. In the angular momentum coupling language, the Dicke states are the $ J = N/2 $ total spin sectors with eigenvalue $ m $, such that we may write
\begin{align}
    |k \rangle \equiv | J = \tfrac{N}{2}, m = k- \tfrac{N}{2} \rangle 
    \label{jmmapping}
\end{align}
where $m\in\{-J,-J{+}1,\dots,J\} $. We may equally define $x$-basis number states by performing a $y$-rotation,
\begin{equation}
\ket{k}^{(x)}=e^{i\pi S^y/4 }\ket{k} .
\label{yrotkx}
\end{equation}
These state are eigenstates of the $ S_x $ operator
\begin{align}
S_x \ket{k}^{(x)} = (2k-N) \ket{k}^{(x)} .
\end{align}

A class of states that are physically preparable under these assumptions are spin coherent states, defined as 
\begin{subequations}\label{eq:init_sc}
\begin{align}
\ket{\alpha,\beta} \rangle
&=\bigotimes_{\ell=1}^{N}\!\bigl(\alpha\ket{0}_\ell+\beta\ket{1}_\ell\bigr),
\label{eq:init_sc:product}\\
&=\frac{(\alpha a^\dagger+\beta b^\dagger)^{N}}{\sqrt{N!}}\ket{0} , \label{eq:init_sc:schwinger}
\end{align}
\end{subequations}
where $ |\alpha|^2+|\beta|^2=1 $.  In the above, we have written both the original qubit formulation and the equivalent Schwinger boson representation \cite{byrnes2024multipartite}.

\subsection{The protocol}
\label{sec:protocol}

We now briefly summarize the quantum repeater protocol of Ref. \cite{pyrkov2025quantum}.   

\subsubsection{Initial state} 

All nodes are initialized to a spin-coherent state polarized along the  $ S_x $  direction (i.e., each initial state of node is  $\ket{+x}^{\otimes N}$, restricted to a symmetric Dicke subspace). Therefore, for a chained system of $M$ ensembles, the global initial state is simply the direct product of the initial states of each node:
\begin{equation}\label{eq:init_global}
\ket{\phi_0}=\bigotimes_{j=1}^{M}\Big\vert \tfrac{1}{\sqrt2},\tfrac{1}{\sqrt2} \Big\rangle\!\Big\rangle_j .
\end{equation}
The spin coherent states can be written in the Dicke basis $ \ket{k} $ for a single ensemble as
\begin{equation}
\ket{+x}^{\otimes N}
=
\Big\vert \tfrac{1}{\sqrt2},\tfrac{1}{\sqrt2} \Big\rangle\!\Big\rangle
=\frac{1}{\sqrt{2^{N}}}  \sum_{k=0}^{N}\sqrt{\binom{N}{k}} \ket{k} .
\label{xpol}
\end{equation}
The Dicke amplitudes are sharply concentrated around $k\approx N/2$ with standard deviation $\sim\sqrt{N}/2$. This amplitude concentration justifies the window truncation and the phase-kick reduction used later. In the angular momentum language (\ref{jmmapping}), for even $ N $, the distribution (\ref{xpol}) is centered at $m=0$ with standard deviation $\sqrt{N}/2$.  For odd $N$, two maxima occur at $m=\pm\tfrac12$.  This does not affect any of the algorithms used in the following.

\subsubsection{Entanglement generation}

The neighboring ensembles interact through an alternating density–density coupling, 
\begin{equation}\label{eq:H_na}
H=\sum_{j=1}^{M-1}(-1)^j\,n_{a,j}\,n_{a,j+1}
\end{equation}
%
which is locally equivalent to an $S_z S_z$ interaction \cite{pyrkov2013entanglement,rosseau2014entanglement,treutlein2006microwave} since
\begin{equation}
n_{a,j}\,n_{a,j+1}=\tfrac14\,({S_{z,j}}+\hat N_j)({S_{z,j+1}}+\hat N_{j+1}) .
\end{equation}
Evolving the Hamiltonian for a time $t$, the initial entangled state is
\begin{align}
    \ket{\Psi(t)}=U(t)\ket{\phi_0}
\end{align}
%
%
The terms within (\ref{eq:H_na}) commute, permitting the decomposition of the unitary into a
\begin{equation}
U(t)=\prod_{j=1}^{M-1}\exp\!\big[-i\,(-1)^j\,n_{a,j}\,n_{a,j+1}\,t \big] 
\label{timeevolop}
\end{equation}
with each gate at two sites contributing only a phase on states 
on sites $ j, j+1 $ since $n_a$ are diagonal for the basis states $\ket{k}$.

The time evolved wavefunction takes a closed form that depends upon whether $M$ is even or odd.  For even $ M $, the wavefunction takes the form 
\begin{multline}\label{evenpsiM}
|\psi_M\rangle=
\frac{1}{\sqrt{2^{\frac{MN}{2}}}}
\sum_{k_1,k_3,\ldots,k_{M-1}=0}^{N}
\left(\prod_{i\in\mathrm{odd}}\sqrt{C_{N}^{k_i}}\right)
\\
\times\,|k_1\rangle_{1}\;
\Big\vert \frac{e^{i(k_{1}-k_{3})t}}{\sqrt{2}},\,\frac{1}{\sqrt{2}}\Big\rangle\!\Big\rangle_{2}\;
|k_3\rangle_{3}\;
\Big\vert \frac{e^{i(k_{3}-k_{5})t}}{\sqrt{2}},\,\frac{1}{\sqrt{2}}\Big\rangle\!\Big\rangle_{4}
\\
\cdots\!\otimes\;
|k_{M-1}\rangle_{M-1}\;
\Big\vert \frac{e^{ik_{M-1}t}}{\sqrt{2}},\,\frac{1}{\sqrt{2}}\Big\rangle\!\Big\rangle_{M}
\end{multline}
Meanwhile for odd $ M$ it is
\begin{multline}\label{oddpsiM}
|\psi_M\rangle=
\frac{1}{\sqrt{2^{\frac{(M+1)N}{2}}}}
\sum_{k_1,k_3,\ldots,k_{M}=0}^{N}
\left(\prod_{i\in\mathrm{odd}}\sqrt{C_{N}^{k_i}}\right)
\\
\times\,|k_1\rangle_{1}\;
\Big\vert \frac{e^{i(k_{1}-k_{3})t}}{\sqrt{2}},\,\frac{1}{\sqrt{2}}\Big\rangle\!\Big\rangle_{2}\;
|k_3\rangle_{3}\;
\Big\vert \frac{e^{i(k_{3}-k_{5})t}}{\sqrt{2}},\,\frac{1}{\sqrt{2}}\Big\rangle\!\Big\rangle_{4}
\\
\cdots\!\otimes\;
\Big\vert \frac{e^{i(k_{M-2}-k_{M})t}}{\sqrt{2}},\,\frac{1}{\sqrt{2}}\Big\rangle\!\Big\rangle_{M-1}\;
|k_{M}\rangle_{M}
\end{multline}
We see that for even $M$, the last site can be written as a spin coherent state, while for odd $M$, the last site must be written in terms of a Dicke number state.

\subsubsection{Measurement}

The next step in the quantum repeater protocol is to  projectively measure all the intermediate nodes in the $x$ basis.  This creates entanglement between the first and last nodes, in a step analogous to entanglement swapping. 
The projection operator corresponding to outcome $q$ at node $j$ is given by
\begin{align}
M_q^{(j)} = \ket{q}^{(x)}_j \bra{q}_j^{(x)} .
\label{eq:measurementop}
\end{align}
%
It is beneficial to introduce a small equatorial offset before measurement. This is accomplished by applying a local phase rotation on the mode $a$,
\begin{equation}
V^{(j)}(\phi)=e^{i n^a_j \phi},
\label{eq:equatorrot}
\end{equation}
applied immediately prior to the measurement.

For the measurement outcomes $\vec q=(q_2,q_3,\ldots,q_{M-1})$ on the intermediate nodes $j=2,\ldots,M{-}1$, the post-selected (unnormalized) state is
\begin{equation}
\ket{\Psi_{\vec q}}=
\left(\bigotimes_{j=2}^{M-1} M_{q_j}^{(j)}\,V^{(j)}(\phi)\right)
\ket{\psi_M} .
\label{eq:projected_state}
\end{equation}
The final unnormalized state after all the measurements is
\begin{equation}
\begin{aligned}
| \Psi_{\vec{q}} \rangle
&= \frac{1}{\sqrt{2^{\frac{MN}{2}}}}
   \sum_{k_1,k_3,\dots,k_{M-1}=0}^{N}
   \sqrt{C_N^{\,k_1}}
   \left(\prod_{j=2}^{M-1}\Omega_{q_j}^{(j)}\right) \\
&\quad\times |k_1\rangle\,
  \Big|\frac{e^{i k_{M-1} t}}{\sqrt{2}},\,\frac{1}{\sqrt{2}}\Big\rangle\!\Big\rangle_{M} ,
  \label{finalbigpsi}
\end{aligned}
\end{equation}
where we defined
\begin{align}
\Omega_{q}^{(j)} = \left\{
\begin{array}{cc}
\langle q |^{(x)} | \frac{e^{i(k_{j-1} - k_{j+1})t + i \phi }}{\sqrt{2}},\frac{1}{\sqrt{2}}\rangle\rangle
 & j \in \text{even} \\
e^{i k_j \phi} \sqrt{ C^{k_j}_{N} } \langle q|^{(x)} | k_j \rangle &  j \in \text{odd}
\end{array}
\right.   .
\end{align}
%
The matrix elements are explicitly given by
\begin{equation}
\langle q |^{(x)} | \frac{e^{i\alpha}}{\sqrt{2}},\frac{1}{\sqrt{2}}\rangle\rangle = i^{N-q} e^{i N  \alpha /2} \sqrt{C_N^{q}}\cos^{q} \frac{\alpha}{2} \sin^{N-q} \frac{\alpha}{2}
\end{equation}
and
\begin{equation}
\begin{aligned}
\langle q|^{(x)}\,|k\rangle
&= \frac{1}{\sqrt{q!\,(N-q)!\,2^N}}
   \sum_{l=0}^{q}\sum_{m=0}^{N-q}
   C_q^{\,l}\,C_{N-q}^{\,m}\,(-1)^{N-q-m} \\
&\times \sqrt{(l+m)!\,(N-l-m)!}\,\delta_{k,l+m}.
\end{aligned}
\end{equation}
where $\delta_{i,j}$ is the Kronecker delta \cite{byrnes2021quantum}. The probability of obtaining this outcome labeled $ \vec{q} $ is
\begin{align}
p_{\vec{q}} = \langle \Psi_{\vec{q}} | \Psi_{\vec{q}} \rangle .
\end{align}


\subsection{Quantifying entanglement}

The long-distance bipartite entanglement of the states can be quantified via the von Neumann entropy
\begin{equation}
E=-\sum^{N}_{l=0} \lambda_l \log_2 \lambda_l ,
\label{entropy}
\end{equation}
where $\lambda_l$ are the eigenvalues of the density matrix
\begin{equation}
\rho_M = \frac{\text{Tr}_1  |\Psi_{\vec{q}} \rangle\langle\Psi_{\vec{q}} |}{p_{\vec{q}}} =
\frac{1}{p_{\vec{q}}} \sum^N_{k_1=0} \langle k_1|\Psi_{\vec{q}} \rangle\langle \Psi_{\vec{q}} |k_1\rangle .
\label{rhom}
\end{equation}
The normalized entanglement $E(t)$ reads as
\begin{equation}
E_{\mathrm{norm}}(t)=\frac{E(t)}{E_{\max}} ,
\label{entnorm}
\end{equation}
where $ E_{\max}=\log_2(N{+}1) $ is the maximum entanglement between two qudits of dimension $ N + 1 $.

\subsection{Decoherence}

One of the strong candidates for physical implementation of our quantum repeater are atomic ensembles in optical cavities.  In Ref. \cite{pyrkov2013entanglement}, a protocol for generating $S_zS_z$ was devised based on a shared photonic mode between two ensembles. 
In the present system, the dominant mechanisms driving decoherence are spontaneous emission and photon loss. \cite{rosseau2014entanglement}. 
This arises from the fact that the shared off-resonant photonic mode possesses a non-zero probability of inducing atomic excitation.
The photons also have the possibility of leakage through the cavities and the connecting optical fiber. At the ensemble level, these effects are well-captured by collective $S_z$ dephasing. Therefore, we may include dephasing during all entangling operations via the master equation
\begin{equation}
\frac{d\rho(t)}{dt}
=-i[H,\rho]
-\frac{\gamma}{2}\sum_{j=1}^{M}\!\left[(S_j^z)^2\rho-2S_j^z\rho S_j^z+\rho(S_j^z)^2\right],
\label{eq:dephasing_master} %
\end{equation}
where $H $ is the entangling Hamiltonian (\ref{eq:H_na}).  This model of decoherence is also appropriate for other methods of generating the $S_zS_z$ interaction, using state dependent forces, for example, proposed in Ref. \cite{kurkjian2013spin}.  In this context $S_z$ dephasing can occur due to technical noise in each of the traps during the evolution to interact the two clouds.

\begin{figure*}[t]
	\centering
	\includegraphics[width=1.8\columnwidth]{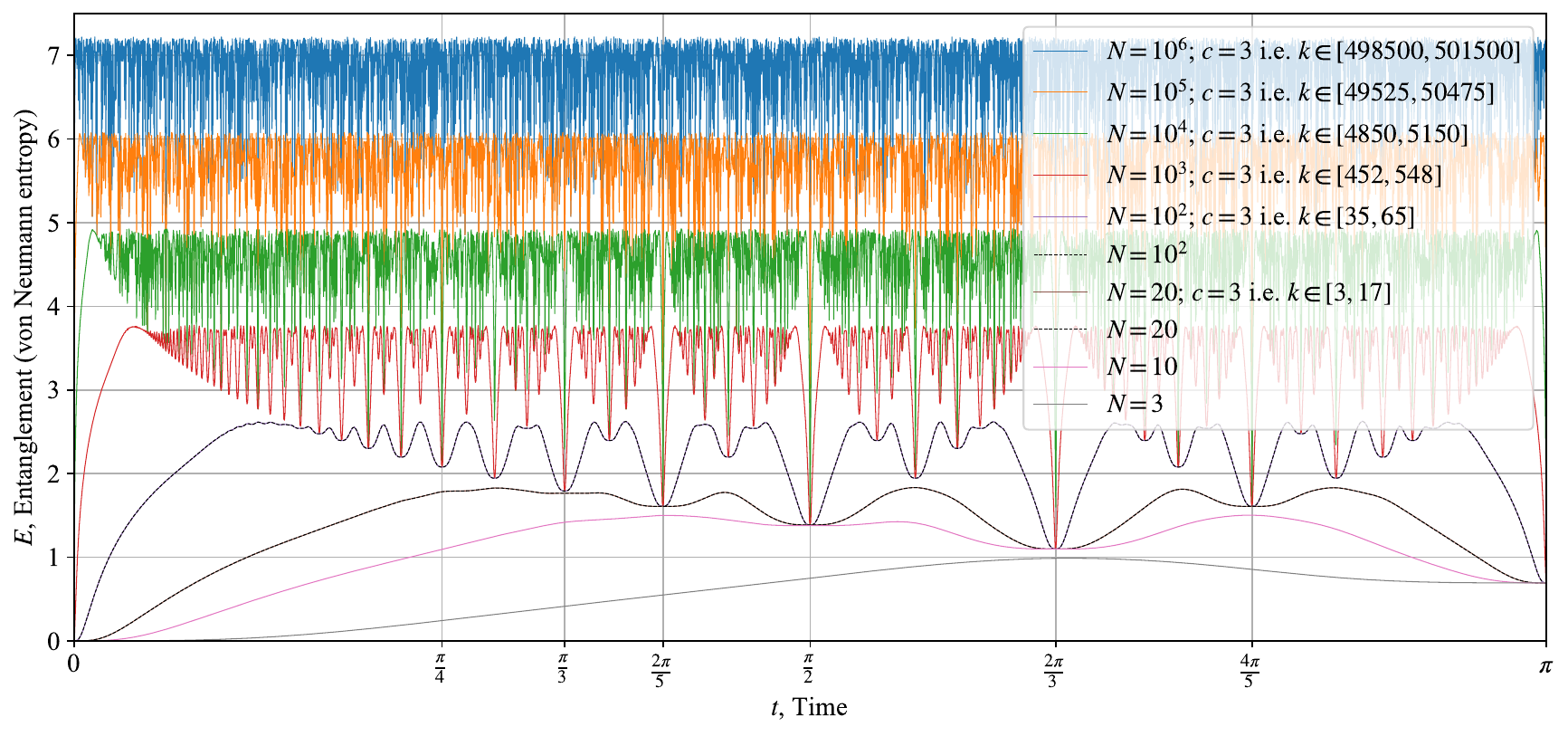}
	\caption{%
  The end--to--end entanglement without normalization (\ref{entropy}) for the chain $M=3$ and the ensemble sizes as marked. The sizes of the truncated window in Fock space for different N are also marked.    
  }
  \label{fig:endtoend-E-over-Emax}
\end{figure*}

\section{Numerical methods}

In this section, we provide a detailed description of the numerical methodologies employed to evaluate the performance of the macroscopic-entanglement-based quantum repeater protocol described in Sec.~\ref{sec2}.

\subsection{Method I: Exact simulation}

We first note that all our calculations throughout this paper are performed in the symmetric subspace of $ N $ qubits, such that only the Dicke states (\ref{eq:dicke:a}) only need to be used.  This reduces the Hilbert space per ensemble from $2^N$ to $ N + 1 $.  

For small scale systems we may perform a full Hilbert space calculation without truncation.   
%
We construct the full many-body Hamiltonian \(H\) explicitly in the Hilbert space of dimension \((N{+}1)^{M}\) and obtain the time-evolution operator (\ref{timeevolop}) followed by the measurements (\ref{eq:measurementop}). Formulas for the basis transformation (\ref{yrotkx}) are available in Ref. \cite{byrnes2021quantum}.  Here $ t $ is fixed to a numerical value with a grid spacing $ \Delta t $ such that the unitary operations can be performed numerically. 
The exponential growth of the Hilbert space limits this exact method to a small range of $N$ and $M$ on standard hardware. Even so, these calculations provide a vital baseline for benchmarking larger-$N$ approximations.
This allow us to pinpoint peak times near magic points, check symmetry relations, and define the exact envelope for $E(t)$.

\subsection{Method II: Fock space truncated window approximation}

For intermediate ensemble sizes, we work in a truncated but otherwise numerically exact Hilbert space.
Locally, we truncate the basis on each node to a finite window centered around the dominant Dicke components, as defined by
\begin{align}
    \frac{N}{2} - K \le k \le  \frac{N}{2} + K 
        \label{bigKdef}
\end{align}
where the half-width of the window is defined as
\begin{equation}
    K = \lfloor \frac{c\sqrt{N}}{2} \rfloor.
\end{equation}
%
In this context, we consider $c$ as a tunable empirical parameter. It is worth to note that $\sqrt{N}/{2}$ is a standard deviation for the Binomial distribution and taking the truncation window of the three standard deviations allows to take into account $99.7\%$ of the largest terms of our quantum state in Fock space.

At the same time, the effective Fock space dimension is then reduced to 
\begin{equation}
    D_{\mathrm{trunc}} \approx (c\sqrt{N}+1)^M .
\end{equation}
It provides a huge numerical efficiency of this method which we discuss in the next section.

Since the fine temporal structure in $E(t)$ emerges on a scale $t \sim 1/N$, a grid spacing 
\begin{equation}
    \Delta t \approx \frac{\pi}{c_t N}
\end{equation}
with $ c_t \gtrsim 100 $ is sufficient to resolve the oscillatory microstructure for all $N$.  






\section{Decoherence-free case}

We first examine the decoherence-free case and show that our methods can extend the simulation range of the quantum repeater protocol to macroscopic ensemble sizes.  We show that the system size can be dramatically increased over Ref. \cite{pyrkov2025quantum} using our numerical methods.

Figure~\ref{fig:endtoend-E-over-Emax} shows $E(\tau)$ for $M=3 $ ensembles and particle numbers $N\in\{3,10,20,100,10^3,10^4,10^5,10^6\}$, computed using exact solution for small N and Method II for large N. 
The outcomes of the measurements on the intermediate nodes equal to $N$ in the $x$-basis. 
The dynamics of the system with time $t$ exhibit a periodicity of  $2\pi$, and its evolutionary behavior in the interval $t\in[\pi,2\pi]$ is the mirror image of that on $t\in[0,\pi]$. 
Therefore, utilizing these symmetries, restricting the time range to $t\in[0,\pi]$ is sufficient to characterize the entire evolutionary process.  


For small $N$, the dynamics are smooth and exhibit a single broad maximum. 
As $N$ increases, the entangled states show a rich structure with fractal characteristics similar to the ``devil's crevasse'' entaglement curve seen in Ref. \cite{byrnes2013fractality}.  The curve takes the form of a slowly varying envelope with a fixed height (the ``entanglement ceiling''), with sharp dips at times that are a rational number multiple of $ \pi $. The fine structure becomes progressively denser with increasing $N$; the typical spacing between adjacent dips is on the order of $1/N$, consistent with the narrowing expected from phase accumulation in large collective spins.  

Our numerics successfully handle much larger qubit numbers up to $N = 10^6$.  The largest system size for which entanglement was computed in Ref. \cite{pyrkov2025quantum} was $ N =20 $, hence this is a considerable improvement over our previous methods.  One important conclusion from calculating larger systems is that it directly shows our protocol successfully generates entanglement even in the macroscopic limit. Entanglement is rapidly established in a timescale  $t \sim 1/N $ and remains at a high and stable level ($E_{\mathrm{norm}} \approx 0.5$) throughout the simulated timescale. This indicates that under the protocol, macroscopic entanglement can indeed be distributed \cite{byrnes2013fractality}. The distributed entanglement is macroscopic in the sense that the amount of entanglement is $ \sim  E_{\max} $.  


We comment that according to popular folklore, the large $ N $ limit of spin ensembles is a type of ``classical'' limit, due to the normalized spin variance tending to zero for increasing $ N $ in spin coherent states.  For example, the normalized spin variance of the $ x $-polarized spin coherent state (\ref{xpol}) is $ \text{Var} (S_z) / \langle S_z \rangle \propto 1/N $, which diminishes to zero for large $ N $.  Our current calculations show that this is not necessarily the case, as even for large $ N $ entanglement is present.  This result shows that the entanglement dynamics generated by our protocol is scalable as the atomic ensemble reaches realistic particle numbers.

\begin{figure}[t]
  \centering
  \includegraphics[width=\linewidth]{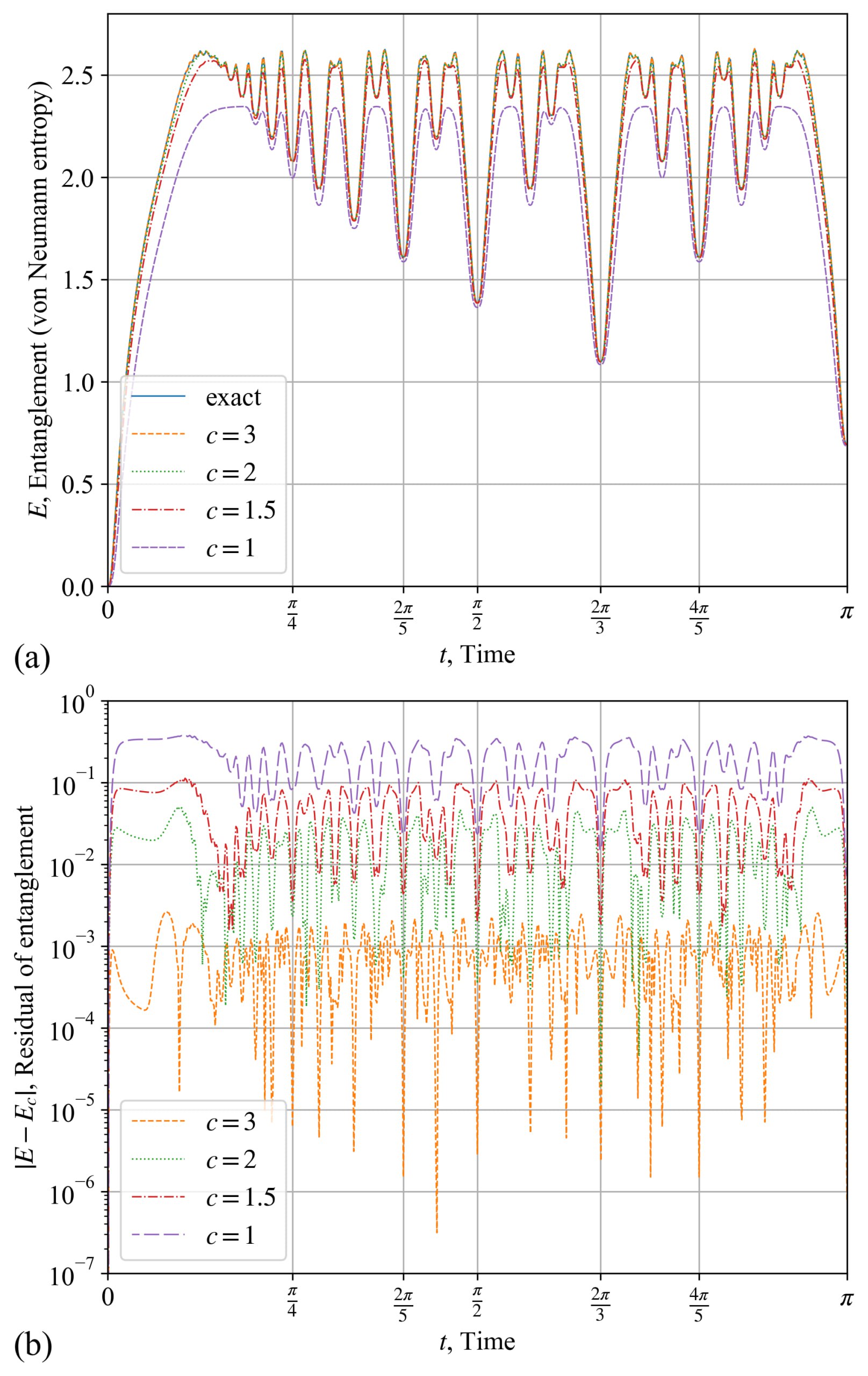}
  \caption{Precision of the entanglement calculation with the Fock space truncated window approximation method  (a) comparison of entanglement for different window sizes with exact solution; 
(b) exact error in entanglement calculations for different truncation window sizes.   %
}
  \label{fig:precision}
\end{figure}

We now examine the precision of the approximation in dependence on the truncated window size.   
Fig.~\ref{fig:endtoend-E-over-Emax} reveals that when we consider the window within 3 standard deviations $c=3$ the end-to-end entanglement calculated with our Fock space truncation window method is undistinguishable from the exact solution by eye.  
Fig. \ref{fig:precision} shows the precision of our truncation window method in dependence on the window size in values of standard deviations for the case of $N=100$. From Fig. \ref{fig:precision}(a) we can see even for $c=1.5 $ precision of our approximation is pretty good by eye and leaves room for calculations with smaller truncation windows and even bigger $N$. Fig. \ref{fig:precision}(b) shows that the precision of our method for the truncated window of 3 standard deviation is of the order of $10^{-3}$ and decreases to $10^{-1}$ for windows of one standard deviation. At the magic times the precision of our approximation is of order of $10^{-5}$ or lower.

\section{Including Decoherence}

\begin{figure}[t]
  \centering
  \includegraphics[width=\linewidth]{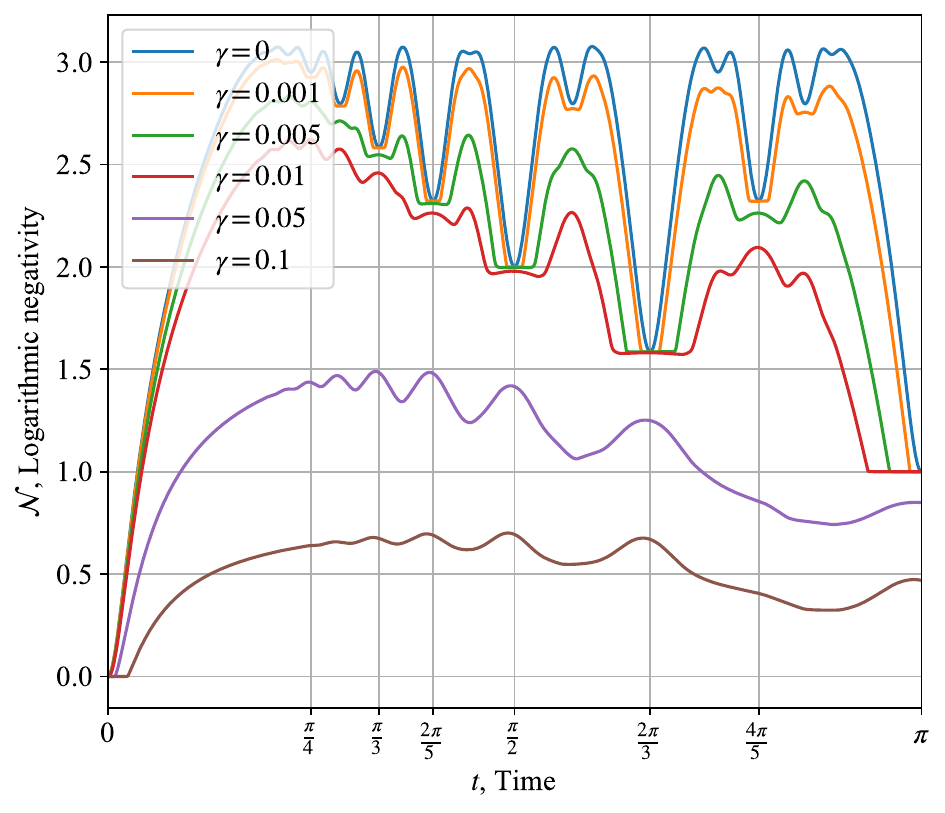}
  \caption{%
  Logarithmic negativity \cal{N} versus time 
  $t$ for $M=3$ and $N=30$ for the dephasing rates as marked.  
  Ensemble dephasing is modeled by the Eq.~(\ref{eq:dephasing_master}).%
  }
  \label{fig:dephasing-scan}
\end{figure}

We now include decoherence into the calculations using our improved numerical methods. We use the dephasing master equation (\ref{eq:dephasing_master}) and the Fock space truncated window method, which allows us to extend the system size to $ N = 30 $. 
While this is smaller than the largest systems examined in the decoherence-free case, it is nevertheless a large improvement over exact methods used in Ref. \cite{pyrkov2025quantum}, which was limited to $ N\le 3$.  
The results are shown in Fig.~\ref{fig:dephasing-scan}.   

We observe that as the dephasing rate $\gamma$ increases, the fine structure is smoothed out and the fractal fluctuations become more suppressed. Under moderate decoherence conditions ($\gamma \leq 10^{-2}$), the overall degradation is insignificant and the envelope generally maintains a monotonically decreasing trend. 
Interestingly, at magic times (e.g. $t = \frac{1}{2} \pi, \frac{2}{3} \pi, \pi, \dots $), the amount of entanglement remains robust with moderate dephasing, where the amount of entanglement being remarkably consistent until a characteristic dephasing rate, beyond which the entanglement starts to diminish. 
In Ref.\cite{pyrkov2025quantum}, it was shown that the width of a single stationary magic-time window is governed primarily by the number of nodes $M$ and broadened as $M$ increases. Consequently, $N$ mainly controls the number of accessible magic times, whereas $M$ mainly controls their temporal robustness, i.e., the window width.
Thus, dephasing primarily reduces contrast without largely  shifting the phase pattern generated by the interacting Hamiltonian.
In short, increasing $\gamma$ suppresses fine peak/valley structure and lowers the achievable upper limit of entanglement; however, as long as decoherence is not excessive, the phase-matching modes of magic time remain quite robust. Accordingly, these traces point to a more experimentally workable range of moderate decoherence, within which end-to-end entanglement can still be generated, and the operative time windows remain clearly discernible and predictable.
With strong dephasing, the entanglement is significantly suppressed.


\begin{figure}[t]
  \centering
  \includegraphics[width=\linewidth]{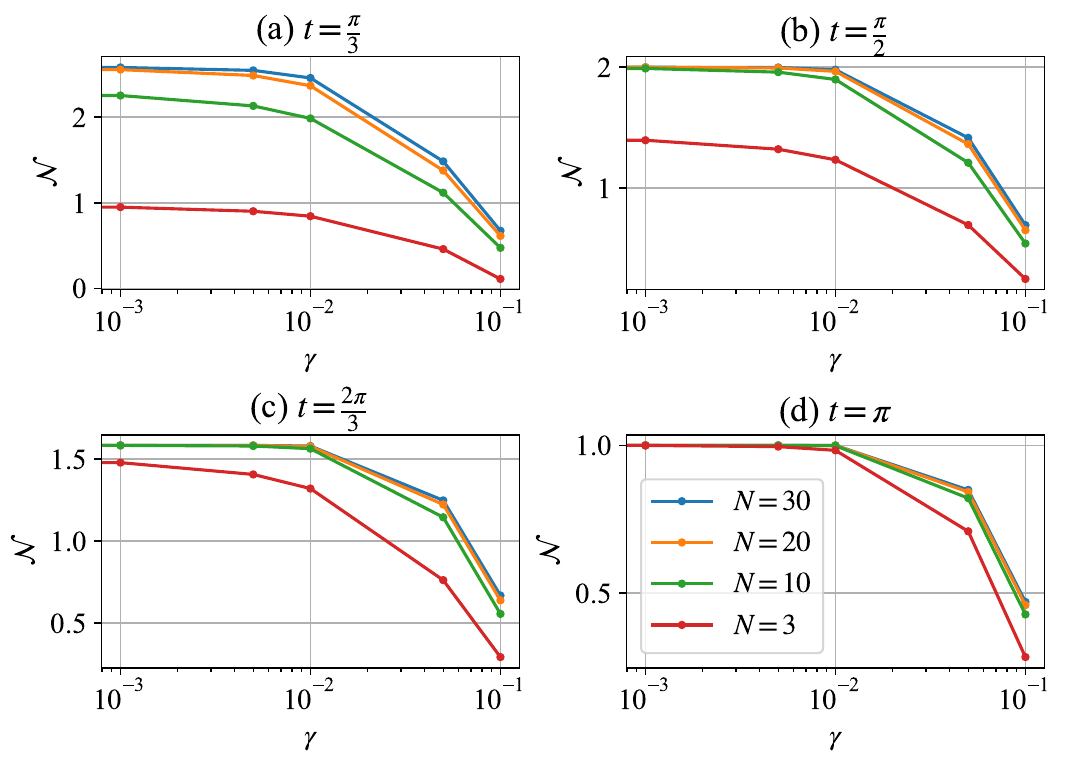}
  \caption{%
Logarithmic negativity $\cal{N}$ plotted as a function of the dephasing rate $\gamma$. The curves correspond to distinct ensemble sizes $N=\{3,10,20,30\}$ and are computed at the four magic times of the ideal protocol.
(a) $t=\pi/3$.  
(b) $t=\pi/2$.  
(c) $t=2\pi/3$.  
(d) $t=\pi$.  
}   \label{fig:magic}
\end{figure}

Figure~\ref{fig:magic} summarizes how dephasing affects the end-to-end entanglement evaluated at four magic times  $t\in \{\pi/3,\;\pi/2,\;2\pi/3,\;\pi\}$. 
For all panels, moderate dephasing $\gamma \lesssim 10^{-2}$ keeps the normalized entanglement $E/E_{\max}$ essentially unchanged, its dependence on $\gamma$ is weak, and near later magic moments (such as $2\pi/3$ and $\pi$), $E/E_{\max}$ almost plateaus within a certain range.
Furthermore,  it can be observed that with increasing evolutionary events, larger atomic numbers exhibited lower sensitivity to the coherence decay factor, demonstrating stronger robustness.
In contrast, for stronger dephasing $\gamma \gtrsim 10^{-2}$, the entanglement is significantly suppressed and the curves compress toward small values, reducing the visibility of size-dependent differences. Regarding finite-size effects, the earliest magic time $t=\pi/3$ shows minimal dependence on $N$, whereas at later magic times, the curves separate modestly, with smaller ensembles exhibiting slightly larger $E/E_{\max}$ and slightly reduced sensitivity to $\gamma$. Finally, the absolute entanglement decreases systematically for all cases shown in Fig.\ref{fig:magic}, consistent with increased dephasing-induced decay at longer interaction times. Overall, these results confirm that the protocol's magic-time entanglement is robust to weak dephasing, in agreement with our observations in  Fig.~\ref{fig:dephasing-scan}.

\section{Summary and Conclusions}

We have investigated a quantum protocol for distributing long-distance macroscopic entanglement on ensembles of qubits.  In the original work of Ref. \cite{pyrkov2025quantum}, exact evaluation methods limited the calculation to relatively small ensemble sizes: for the decoherence-free case the calculations were limited to $ N \le20 $, and  $ N \le 3 $ after including decoherence.  In this work, we were able to extend the calculation regimes to $ N \le 10^6 $ for the decoherence-free case and $ N \le 30 $ for the decoherence case.  This was made possible by controlled truncation methods that retains the accuracy of the calculations despite reducing the Hilbert space size.  

Our calculations have confirmed that the approach of Ref. \cite{pyrkov2025quantum} should indeed be viable in the macroscopic limit.  The dependence of the entanglement follows a ``devil's crevasse'' characteristic form, similar to that seen in Ref. \cite{byrnes2013fractality}.  This originates from the $ S_z S_z $ interactions that are used as the fundamental elements for generating entanglement between two nodes.  We observed that as the ensemble size $N$ increases, an interference pattern appears on top of a slowly varying envelope whose height (the ``entanglement ceiling'') that is only weakly dependent on $N$.  The fine structure exhibits an increase in density with larger $N$, where the typical interval between successive depressions is approximately $1/N$.
An increase in the parameter $\gamma$ results in the attenuation of the fine structure and a reduction in the upper limit. However, when degradation is moderate, the magic-time pattern remains comparatively robust. 

Our key conclusions that result from this study is as follows.  First, in the decoherence-free case, the amount of entanglement that is generated is typically $ \sim E_{\max}/2 = \frac{\log_2 (N+1) }{2} $.  This is the same order as the maximal amount of entanglement that can be attained between two completely symmetric ensembles with $ N $ qubits.  This increases with $ N $, and hence we consider that in this limit macroscopic entanglement distribution can be attained. It is important to note that the our quantum repeater protocol does not include any dependence on $ N $ since all operations are symmetric under local qubit interchange.  This make the protocol highly efficient in generating macroscopic entanglement. Second, the entanglement that is distributed at magic times tends to be of a robust nature and only has a weak dependence with the dephasing rate.  This is a fortunate coincidence as the magic times also correspond to times where the quantum repeater protocol do not degrade with $M $ \cite{pyrkov2025quantum}.

For future work, one direction is to consider other types of elementary entanglement generation between neighboring nodes. An alternative method use quantum nondemolition (QND) measurements, which have shown to be able to produce maximal entanglement between qubit ensembles \cite{chaudhary2023macroscopic}.  One challenge to be overcome with this approach is that the maximal entanglement is attained using the special properties of two-ensemble entanglement, which allows one to overcome the stochastic nature of the QND entanglement generation.  Similar methods for $M$ ensembles would be needed to make it a deterministic scheme.  Another direction is to consider applications of such remotely distributed macroscopic entanglement, towards distributed quantum computing \cite{byrnes2012macroscopic} or quantum sensing  \cite{fadel2023multiparameter}.


\FloatBarrier

\begin{acknowledgments}

This work is supported by the SMEC Scientific Research Innovation Project (2023ZKZD55); the National Natural Science Foundation of China (92576102); the Science and Technology Commission of Shanghai Municipality (22ZR1444600); the NYU Shanghai Boost Fund; the China Foreign Experts Program (G2021013002L); the NYU-ECNU Institute of Physics at NYU Shanghai; the NYU Shanghai Major-Grants Seed Fund; and Tamkeen under the NYU Abu Dhabi Research Institute grant CG008. I.D.L. and A.N.P. are supported by the state task, State Registration No. 124013000760-0.
\end{acknowledgments}

 \section*{Conflict of Interest}

 The authors declare no conflict of interest.

 \section*{data availability}

 Data supporting the findings of this study are available from the corresponding author upon request.

\bibliographystyle{apsrev4-2}
\bibliography{ref}

\end{document}